# The evolution of magnetic tower jets in the laboratory

A. Ciardi[1,2], S.V. Lebedev[2], A. Frank[3,4], E.G. Blackman[3,4], J.P. Chittenden[2], C.J. Jennings[6], D.J.Ampleford[6], S. N. Bland[2], S. C. Bott[2], J. Rapley[2], G. N. Hall[2], F. A. Suzuki-Vidal[2], A. Marocchino[2], T. Lery[5] and C. Stehle[1]

[1]*LUTH, Observatoire de Paris et UMR 8102 du CNRS, 92195 Meudon , France.*

[2]*The Blackett Laboratory, Imperial College, London SW7 2BW, UK*

[3]*Department of Physics and Astronomy, University of Rochester, Rochester NY USA*

[4]*Laboratory for Laser Energetics, University of Rochester, Rochester NY USA*

[5]*Dublin Institute for Advanced Studies, Dublin, Ireland*

[6]*Sandia National Laboratory, Albuquerque, New Mexico, USA*

**Abstract**

The evolution of laboratory produced magnetic jets is followed numerically through three-dimensional, non-ideal magnetohydrodynamic simulations. The experiments are designed to study the interaction of a purely toroidal field with an extended plasma background medium. The system is observed to evolve into a structure consisting of an approximately cylindrical magnetic cavity with an embedded magnetically confined jet on its axis. The supersonic expansion produces a shell of swept-up shocked plasma which surrounds and partially confines the magnetic tower. Currents initially flow along the walls of the cavity and in the jet but the development of current-driven instabilities leads to the disruption of the jet and a re-arrangement of the field and currents. The top of the cavity breaks-up and a well collimated, radiatively cooled, "clumpy" jet emerges from the system.

# I. INTRODUCTION

In recent years a number of scaled, high energy density experiments were devised to investigate a wide range of complex dynamical flows and processes of relevance to astrophysical problems. The experimental approach can in general advance not only our understanding of the underlying physics but may also provides, in addition to astronomical observations, stringent tests to numerical codes and models. The problems tackled in the laboratory range from core-collapse supernovae and photoevaporated molecular clouds to the studies of radiative shock waves and equations of state of planetary interiors (see [1] for a review). An issue that has attracted some attention in the laboratory astrophysics context is the formation and propagation of jets. Using lasers[2,3,4,5] and pulsed-power facilities[6,7,8,9,10], the jets were produced by purely hydrodynamic means and relied in general on strong radiative cooling. However, there is now a general consensus that the jet formation mechanism at play in astronomical objects is not hydrodynamic but magneto-hydrodynamic, and magnetic fields are an essential component that needs to be added to the laboratory jets production mechanism.

The presence of jets is indeed ubiquitous in space. The astronomical objects with associated jets range from super-massive black holes present in the centre of active galactic nuclei (AGN) (see for example[11]) to young star[12]. In addition, gamma ray bursts (GRB) (see [13] for a review) and supernovae may also have jets associated with them[14,15,16,17,18,19,20]. In general, jets are thought to be powered by the combination of rotation and magnetic fields, which extract the rotational energy from an accreting system and create magnetic stresses which accelerate and collimate the flow[21,22,23,24,25,26,27,28,29,30,31,32,33]. Depending on the details of the models, the winding of an initially poloidal (in the *r-z* plane) magnetic field results in a flow pattern dominated by a toroidal field[34]. A similar situation is also attained when the foot-points of a field line, connecting the disc to a central compact object or connecting different parts of a disc, rotate with different angular velocities. In such cases, the relative angular displacement of the foot-points causes one of the foot-points to move ahead of the other and the field loop to twist. The induced toroidal component results in an increase of the magnetic pressure which drives the expansion of the loop itself[35]. In the magnetic tower scenario[36,37], the outcome is a magnetic cavity consisting of a highly wound up toroidal field which accelerates the flow. In this case, the presence of an external plasma medium was shown to be necessary to confine the magnetic cavity, which would otherwise splay out to infinity within a few rotations[36]. The basic picture of magnetic tower evolution has also been confirmed numerically by several authors[38,39,40,41,42], and it is in this context we have develop a series of laboratory astrophysics experiments to study the jet formation mechanism through the interaction of a purely toroidal field with an extended plasma background medium. Both experiments[43] and numerical simulations[44,45] showed the formation of a magnetic cavity consisting of toroidal field loops with a jet embedded on its axis and due to the supersonic expansion, a shock envelope around the magnetic cavity. Indeed, recent work on astrophysical magnetic towers[46,47] was extended to include the supersonic propagation of the magnetic cavity, which as observed in the experiments, results in a shock envelope around the magnetic tower.

The present paper reports on three-dimensional (3D) simulations of laboratory jets, extending on our previous two-dimensional simulations[44] and allowing in particular the study of the late stages of the magnetic tower evolution, when non-axisymmetric effects become important. The first part of the paper describes the model and initial conditions. The general dynamics of the laboratory jets are then discussed followed by a more detailed description of the various stages of jet evolution. We conclude with a summary of the results and discussion.

## II. THE MODEL

We perform the numerical simulations with an explicit, parallel version of the 3D resistive magneto-hydrodynamic (MHD) code GORGON[48]. The MHD equations are solved on a three-dimensional Cartesian grid in the single fluid approximation. However, the ion and electron components of the plasma are allowed to be out of thermodynamic equilibrium with respect to each other and their relative energy equations are solved separately. The equations of conservation of mass, momentum and internal energy are given by:

$$\frac{\partial \rho}{\partial t} + \nabla \cdot (\rho \mathbf{v}) = 0 \tag{1}$$

$$\frac{\partial}{\partial t}(\rho \mathbf{v}) + \nabla \cdot (\rho \mathbf{v} \mathbf{v}) = -\nabla(p_i + p_e) + \mathbf{j} \times \mathbf{B} \tag{2}$$

$$\frac{\partial \varepsilon_i}{\partial t} + \nabla \cdot (\varepsilon_i \mathbf{v}) = -p_i \nabla \cdot \mathbf{v} - \nabla \cdot \mathbf{q}_i + \Delta_{ie} \tag{3}$$

$$\frac{\partial \varepsilon_e}{\partial t} + \nabla \cdot (\varepsilon_e \mathbf{v}) = -p_e \nabla \cdot \mathbf{v} - \nabla \cdot \mathbf{q}_e + \eta |\mathbf{j}|^2 - \Lambda + \Delta_{ei} \tag{4}$$

where $\rho$ is the density and $\mathbf{v}$ is the velocity. The internal energy of the ions $\varepsilon_i$ and the pressure $p_i$ are related by an equation of state of the form $\varepsilon_i = \frac{p_i}{\gamma - 1}$. For the electrons the internal energy $\varepsilon_e$ is given by $\varepsilon_e = \frac{p_e}{\gamma - 1} + Q(\bar{Z})$, where $Q(\bar{Z})$ is the ionization potential energy and depends on the average ionization charge $\bar{Z}$ of the plasma. The latter is calculated from an average-ion Thomas-Fermi model. The adiabatic index is taken to be $\gamma = 5/3$. Radiation effects are included through an optically-thin radiation losses sink term $\Lambda$ and Ohmic heating is given by the source term $\eta|\mathbf{j}|^2$, where $\eta$ is the resistivity. The ion and electron thermal fluxes are given respectively by $\mathbf{q}_i = -\kappa_i \nabla T_i$ and $\mathbf{q}_e = -\kappa_e \nabla T_e$, where $\kappa$ is the thermal conductivity; Braginskii-like transport coefficients are used for $\kappa$ and $\eta$. Finally, $\Delta_{ie}$ and $\Delta_{ei}$ are the energy exchange rates between ions and electrons ($\Delta_{ei} = -\Delta_{ie}$) and couple the two energy equations. The evolution of the electromagnetic fields is followed through the vector potential $\mathbf{A}$. The magnetic induction $\mathbf{B}$ is then only calculated as a diagnostic by $\mathbf{B} = \nabla \times \mathbf{A}$, thus ensuring the constraint $\nabla \cdot \mathbf{B} = 0$ is preserved. In Z-pinch experiments it is often the case, at least initially, that a large fraction of the experimental volume is in vacuum. In the code, the computational "vacuum" is defined for cells with a density below $\rho_{vacuum} = 10^{-7}$ g cm$^{-3}$ and only the vacuum form of Maxwell's equations is solved there; to propagate the fields in the computational vacuum the displacement current in Ampere's law is thus retained $\nabla \times \mathbf{B} = \mu_0 \mathbf{j} + \frac{1}{c^2}\frac{\partial \mathbf{E}}{\partial t}$. Finally, the combinations of Ohm's and Faraday's laws give for the electric field $\mathbf{E} = -\frac{\partial \mathbf{A}}{\partial t} = \eta \mathbf{j} - \mathbf{v} \times \mathbf{B}$ and the electromagnetic fields are then advanced in time by solving:

$$\frac{\partial^2 \mathbf{A}}{\partial t^2} = -c^2 \nabla \times \nabla \times \mathbf{A} - \frac{\mu_0 c^2}{\eta}\frac{\partial \mathbf{A}}{\partial t} + \frac{\mu_0 c^2}{\eta} \mathbf{v} \times \nabla \times \mathbf{A} \qquad (5)$$

Free flow boundary conditions are implemented on all sides of the computational box. An approximation of the experimental current is imposed on the system by appropriately setting the field on the boundaries below the outer electrode. The wires are initiated as a high density gas with a temperature $T = 0.125$ eV; although the transition from solid wires to plasma is not modelled in the code, the expected level of wire ablation rate[49] is nevertheless recovered by this artificial initialization. For the present work the regions representing the electrodes are considered to be thermally insulated and have a high electrical conductivity ($\eta \sim 10^{-6}\,\Omega\,\text{m}$), which ensures a resolvable current layer and appropriate boundary conditions for the fields. However this does not realistically model the electrodes, their ablation and any effects they have on contact with the wires.

The schematic of a radial wire array is shown in Figure 1. The typical set-up consists of 16 tungsten wires of 13 $\mu$m diameter. The wires connect radially two concentric electrodes and the load is then driven on the MAGPIE generator[50], which delivers 1 MA current over 240 ns. The inner and outer electrodes have radii of 2 mm and 20 mm respectively. The formation and evolution of the magnetic tower takes place over the central region above the inner electrode and for computational efficiency the radius of the outer electrode in the simulations is taken to be 12.5 mm. The global magnetic field $\mathbf{B}_G$, indicated in Figure 1, is produced by the current through the central electrode and dominates, at least in the central volume of the array, over the "private" field produced by each wire. Therefore below the plane of the wires $|\mathbf{B}_G|$ takes approximately the standard vacuum value (within ~1-2%) and the field is practically toroidal, $\mathbf{B}_G \cong -\dfrac{\mu_0 I(t)}{2\pi r}\hat{\phi}$, where $I(t)$ is the time-dependent current and $r$ is the radial distance from the axis of the arrays.

## III.  MAGNETIC TOWERS IN THE LABORATORY

### A. The typical evolution of a laboratory jet

In this section we present a general overview of the dynamics of laboratory magnetic jets. Figure 2a shows a time sequence of mass density *x-z* slices while the overall 3D structure is shown in Figure 2b. The sustained ablation of the wires proceeds until the mass in the wire cores is exhausted. Near the inner electrode, where the mass ablation rate is highest, sections of the cores disappear and at ~ 180 ns the formation of the magnetic bubble begins (first panel of Figure 2a). The sudden release of magnetic energy drives a shock in the background plasma, the magnetic field acts as a "piston" sweeping the plasma into a shock-layer enveloping the magnetic cavity. On the axis of the magnetic bubble, the "snowploughed" plasma is compressed and accelerated into a jet (200 ns). Despite the azimuthal modulation present in the background plasma and in the shock envelope, the magnetic tower maintains a very high symmetry. The schematic picture of the magnetic tower that has evolved thus far is given in Figure 3. The current flows around the envelope of the magnetic cavity and through the central jet, which is "pinched" by the toroidal magnetic field present inside the cavity. Radial and axial profiles at 220 ns are shown in Figure 4a and 4b at different heights and radii. The markedly different physical properties present in the jet, magnetic cavity and in the background plasma are evident in the images. The jet is susceptible to the development of current-driven instabilities and both "sausage" and "kink" modes appear in the body of the jet. The growth of large amplitude perturbations and further expansion of the cavity ultimately lead to the detachment of the jet and the break-up of the cavity (240 ns). However, a clumpy jet still emerges from the cavity (270 ns), transporting in a narrow channel mass and energy away from the formation region.

Simulated and experimental self-emission x-ray emission images are shown in Figure 5. The synthetic images are obtained by integration of the plasma emission along either the *x* or *y*-direction. Only optically thin radiation losses are considered and neither opacity effects nor the re-emission of absorbed radiation are taken into account. Although the synthetic images simply serve for a qualitative comparison, the emission structures seen in the experiments are well reproduced. Visible in the images is the early development of perturbations in the jet body and the more pronounced kink-like structures that develop at later times. The emission from the magnetic tower is dominated at all times by the jet. There is also relatively strong emission at the head of the magnetic tower where a bow shock propagates into the ambient medium. In the context of stellar explosion driven by magnetic towers[47], the shocked plasma feeds a "cocoon" of high-pressure material that further confines the magnetic tower. In the present case however, cooling in the shocked plasma reduces significantly the pressure and relatively cold, dense region of material forms at the head of the magnetic tower. Finally, the shock envelope produces in general weaker emission, especially in the lower part of the cavity where the lateral expansion is relatively slow and shock-heating negligible. However, the discrete emission features, corresponding to the density modulation visible in the 3D plots, are also present in the x-ray images.

### B. Initial plasma dynamics and the formation of an ambient medium

The presence of an initial background plasma medium is essential to the evolution of the magnetic tower: it provides mass to the central jet, confinement for the magnetic field and supports the currents circulating in and around the cavity.

Wire ablation proceeds as energy is deposited in the wire through Ohmic heating and to a much lesser degree through thermal conduction. Very early in the current pulse (~ 20 ns) the wires develop a two-component structure consisting of a dense, cold and high-resitivity core surrounded by a relatively hot (few eV), low-resistivity "coronal" plasma[51]. Because of the marked differences in the resistivity, currents preferentially flow in the "coronal" plasma which is accelerated in the axial direction by the $\mathbf{j} \times \mathbf{B}_G$ force. The cold wire cores instead are stationary and during the current pulse act as a continuous source of plasma, feeding the flow above the wires and producing a background plasma environment. Figures 6 and Figure 7 shows the averaged profiles for several quantities of interest at different times. In all cases a toroidal average is performed over $2\pi$. The radial profiles are averaged in the $z$-direction from $z = 5.2$ mm up to the height of the magnetic tower. As seen from Figure 4a, there is little variation in the radial profiles with height, and little information is lost when performing an axial average. On the other hand, the axial profiles are radially averaged from $r = 0$ up to the radial extent of the magnetic cavity but not including the shock layer. In this case, the distinction between the jet and the cavity is lost and what is shown is an average property similar to an unresolved observation. The plots show the electron temperature $T_e$, axial velocity $V_z$, mass density $\rho$, and three dimensionless parameters: the ratio of the toroidal to the axial magnetic field $B_\phi / B_z$, the plasma-$\beta$ and the fast-magnetosonic Mach number $M = |\mathbf{v}| / \sqrt{V_A^2 + C_S^2}$, where $V_A$ and $C_S$ are the Alfven and sound speed respectively.

Above the plane of the wires, the flow from the wires merges and collides, resulting in an increase with height of the temperature (and average ionization) of the plasma (see for example Figure 6). The axial velocity also increases with $z$, although it is due to the Lorentz force acceleration-length extending a few millimetres above the wires. Close to the wires, the resistive diffusion dominates over the transport of the magnetic field and the magnetic field remains in the proximity of the wires.

In general the ambient plasma has a radial density distribution given approximately by $\rho \sim r^{-2}$. Such form is consistent with the density distribution discusses in [43] and we find good agreement (within ~ 10-15%) with the simulated density profiles, at least for heights which are neither too close nor too far from the wires. However, because of the azimuthal modulation present in the flow, the simulated density profiles shows a steeper dependence on $z$ than expected, $\rho \sim z^{7/2}$.

### C. The magnetic cavity

Conditions in the cavity are extremely dynamical and far from homogeneous, large gradients in the temperature, velocity and density are present and currents can randomly flow across the plasma connecting the cavity walls and the jet. The formation begins at ~ 180 ns and it is visible as a localized increase in the temperature $T_e$ and axial velocity $V_z$, (see Figure 7, at $r \sim 2$ mm and at $z \sim 5 - 7$ mm). As the cavity grows in height and expands laterally, the magnetic piston sweeps away the plasma and the average density decreases; with the exception of the axis where the jet forms and the density increases. At the same time there is a considerable increase in the average temperature and velocity, resulting from the conversion of magnetic energy into kinetic energy. The field can increase the kinetic energy of the plasma at a rate given by $\mathbf{v} \cdot (\mathbf{j} \times \mathbf{B})$ and on average $\mathbf{v} \cdot (\mathbf{j} \times \mathbf{B}) > 0$ inside the cavity. Part of the acquired kinetic energy then heats the plasma, either through compressional heating or by irreversible thermalization in shocks, raising the temperature and average ionization. However, because of the low densities, radiation losses cannot efficiently remove the excess of energy and the temperature remains high. Despite the high temperatures, the plasma $\beta$ (defined as the ratio of the thermal to the magnetic pressure) in the cavity is always low and the magnetic pressure dominates over the thermal pressure. In addition to the kinetic energy, the magnetic field can be converted in internal energy of the plasma through Ohmic heating $\eta |\mathbf{j}|^2$. Between ~

180 ns and ~ 215 ns the ratio $\Theta = \frac{\mathbf{v} \cdot (\mathbf{j} \times \mathbf{B})}{\eta |\mathbf{j}|^2}$ increases from $\Theta \sim 3$ to $\Theta \sim 15$, a corresponding increase in the total kinetic energy inside the cavity can be seen in Figure 10). However from ~ 220 ns, with the development of instabilities and steep gradients in the magnetic field, the Ohmic dissipation begins to dominate and $\Theta \sim 1$. We note that both $\mathbf{v} \cdot (\mathbf{j} \times \mathbf{B})$ and $\eta |\mathbf{j}|^2$ are ~ 10 times greater in the jet than in the cavity.

The ratio $B_\phi / B_z$ shows that the field inside the cavity is mainly toroidal and only at late times, with the development of instabilities, the ratio is seen to decrease (see later). Additionally, although much less than $B_\phi$, there is initially a radial component of the field $B_r$ produced by the asymmetries present in the axially converging shock.

The low densities, high temperatures and high velocities present in the magnetic cavity raise the question of whether the assumption of collisionality of the plasma and thus the MHD validity in some regions may be suitable. Estimates of the upper limit of the ion mean-free-path, using the vacuum cut-off density employed in the simulations, shows that the scattering length $\lambda_{ii} \sim 200$ mm is indeed larger than the characteristic size of the system, $D \sim 5$ mm; however the ions are strongly magnetized and the ion Larmor radius is generally small $r_{L_i} \sim 0.3$ mm. Following the discussions of collisionless shock experiments[52] and similarity criteria[53] we define the parameter $K$ as the ratio of the Larmor radius $r_{L_i}$ to the characteristic size $D$ of the system $K = \frac{r_{L_i}}{D} \approx 1.4 \times 10^{-4} \frac{\sqrt{EA}}{D\bar{Z}B}$, where $A$ is the atomic number, $\bar{Z}$ is the average ion charge, $B$ is the magnitude of the magnetic field and $E$ is the ion kinetic energy.

The flow in the cavity is in general super-fast-magnetosonic ($M > 1$) and we use the non-thermal kinetic energy of the ions $E \sim 230 \times 10^3$ eV, corresponding to a characteristic flow velocity $v \sim 5 \times 10^5$ m s$^{-1}$, to obtain an upper estimate of the parameter $K$. Taking typical values ($A \sim 184$, $D \sim 0.005$ m, $\bar{Z} \sim 30 - 60$ and $B \sim 50$ T) gives $K < 0.1$, indicating that the ions in the cavity are well localized and fluid-like behaviour is a good approximation[53]. The estimate of the Larmor radius is however larger than the resolution used in the computations (100–200 μm) and we expect the fluid approximation to break down in some restricted computational cells; nevertheless the very low density plasma regions ($\rho \sim \rho_{vacuum}$) where $K > 1$ play no significant dynamical role. We also performed simulations which included an anomalous resistivity correction term[54], which mostly affects the low density plasma regions. The results showed that while the physical properties of the low density plasma in the cavity can be significantly altered, the overall dynamical evolution of the magnetic tower is not affected. However, important effects such as particle acceleration may still take place. Particularly at the base of the cavity, where following the break-up of the jet, large electric fields and reconnection events may impart enough energy to the ions so that $K > 1$.

In other regions of the magnetic tower the ion scattering length is much smaller than the Larmor radius $\lambda_{ii} \ll r_{L_i}$, the ions are not magnetized and the system is highly collisional, $\frac{\lambda_{ii}}{D} \ll 1$. This contrasts with the astrophysical jets, where the ions are generally well magnetized and weakly collisional. However, the parameter $K$ in the astrophysical context is very small $K < 10^{-6}$ and the presence of plasma micro-fluctuations and magnetic entanglement are further assumed to localize the system along the field lines[53]. Therefore, both the astrophysics and experimental jets can often be studied using fluid-like equations, albeit for different reasons.

### D. The formation and launching of the jet

The jet evolution can be roughly divided into three broad phases. During the first, lasting form ~180 ns to ~ 210 ns, the jet forms inside the cavity under the action of the Lorentz forces which first accelerate the plasma towards the axis and then confine it there. From ~ 210 ns to ~ 240 ns, current-driven instabilities grow in the jet which begins to disrupt; finally in the third phase, at times greater ~ 240 ns, the jet breaks-up and emerges from the cavity.

The density increase on axis during the first phase is evident in the radial profiles for $r \leq 1$ mm. At 210 ns the accumulated mass in a cylinder with a 1 mm radius and 6 mm in height is $\sim 4.1 \times 10^{-5}$ g. This is approximately 3 times the mass present in same volume at 180 ns and it is comparable to the mass coming from the 1 mm gaps produced in the wires when they fully ablate. The plasma beta is initially high ($\beta \geq 1$) in the jet, but decreases as the current is confined to a smaller radius and the magnetic pressure increases. Under the assumption of approximate Bennett equilibrium and considering that ~ 75% of the total current (970 kA at 210 ns) flows within a radius ~ 1 mm, results in a Bennett temperature of ~ 176 eV, comparable to the experimental estimates[43]. In general, the high rate of radiation losses efficiently removes energy from the dense jet and results in temperatures considerably lower than in the surrounding cavity. Above $z \geq 18$ mm the emission of the jet dies out, the flow having efficiently cooled down through radiation losses. We estimate the cooling parameter $\chi$ in the jet as $\chi = \dfrac{\varepsilon}{\tau_H P_R}$, where $P_R$ is the energy lost to radiation radiated per unit time, $\varepsilon$ is the energy density and $\tau_H$ is a characteristic hydrodynamic time, taken to be ~ 50-100 ns. In a radiatively efficient regime $\chi < 1$ and cooling through radiation is important in the dynamics and energy balance of the system. The jet is indeed in this regime, with $\chi \sim 10^{-2}$, while in the cavity $\chi \sim 1-10$. The plasma in the jet is not only compressed and confined, but it is also accelerated by the magnetic field. The characteristic velocities in the jet are higher than those present in the background plasma prior to the passage of the magnetic cavity. This is evident for example in Figure 4, the velocities in jet axis and in the cavity are clearly higher than that present in the background plasma ($z \geq 16$ mm). Overall, the axial velocity of the head of the magnetic tower, $V_{Tower}$, increases linearly with time from $V_z \sim 200$ km s$^{-1}$ at 200 ns to $V_z \sim 800$ km s$^{-1}$ at 250 ns. Such behaviour is well reproduced by assuming that across the shock at the head the tower $P_{mag} = \rho_{bg} \left[ V_{shock} - V_{bg} \right]^2$, where $P_{mag}$ is the average magnetic pressure in the tower, $\rho_{bg}$ and $V_{bg}$ are the density and axial velocity of the background plasma just ahead of the shock, and $V_{shock}$ is the shock axial velocity. Because of the optically thin radiation losses there is no simple relation between piston velocity $V_{Tower}$ and shock velocity $V_{shock}$. However the presence of strong radiation cooling can increases the compression factor well above the adiabatic case and for very high compressions $V_{Tower} \approx V_{shock}$. The velocity $V_{Tower}$ is then given by $V_{Tower} = \sqrt{\frac{P_{mag}}{\rho_{bg}}} + V_{bg}$. Using the simulated values for $P_{mag}$, $\rho_{bg}$, $V_{bg}$ recovers well the temporal variation of the tower velocity $V_{Tower}$. A similar expansion velocity is expected in models of stellar explosions driven by magnetic towers[47].

The initial jet configuration, a current-carrying column of plasma confined by a toroidal field, is known to be prone to disruptive, current-driven instabilities. The growth time of the $m=0$ and $m=1$ modes in the laboratory jet was estimated to be of the order of a few nanoseconds[43], much shorter than the characteristic jet evolution time. Indeed the presence of these modes was seen from the early stages of jet formation[43] and it is reproduced in the simulations presented here. However the resolution that can be afforded in the jet region precludes an in-depth analysis of these instabilities and possible stabilising effects, such the presence of an axial field, which are thought to be present in astrophysical jets[55,56]. Thus in the global simulations discussed here only a qualitative picture is given. The Kruskal-Shafranov criterion gives a semi-quantitative condition for the development of current-driven instabilities, which states that the magnetic twist $\Phi = LB_\phi/rB_z$, where $L$ is the length of the column and $\Phi$ is evaluated at the column radius, should be $\Phi < 2\pi$ for stable configurations. Depending on the details of the initial equilibrium, the stability threshold may be somewhat higher. However for the laboratory jets, where $B_z \ll B_\phi$, the magnetic twist is very large $\Phi \sim 10^3$ and all axial wavelengths above $\sim 200$ $\mu$m are expected to be unstable. At 200 ns the jet shows the presence of many wavelengths with the $m=0$ mode initially dominant. Axial velocity variation along the jet body, due to non-uniform acceleration may also contribute to sausage-like modulations. However, after $\sim 220$ ns the $m=1$ mode dominates and the presence of well developed "kinks" in the jet becomes clearly visible. Twisting of the currents and magnetic field, turn the toroidal magnetic flux into poloidal (in the *r-z* plane) flux and leads to a re-arrangement of the magnetic field. This effect can be seen for example in the Figure 6 and 7 as a decrease over time of the ratio $B_\phi/B_z$. A three-dimensional view of some magnetic field lines is shown in Figure 9 at a time of 230 ns. A toroidal component is still present inside the cavity and in the jet, especially at larger radii; however on the (kinked) jet axis the poloidal component $\mathbf{B}_P$ has become dominant. At the particular time shown the ratio $\frac{B_\phi}{|\mathbf{B}_P|} \leq 1$ for $r \leq 1$ mm, where $|\mathbf{B}_P| = \sqrt{B_r^2 + B_z^2}$.

The effective detachment of the jet and the opening-up of the cavity occurs after ~240 ns, a large amplitude kink disrupt the base of the jet body while the top part of the jet begins to break through the envelope of the magnetic cavity. Axial profiles are shown in Figure 8 at 270 ns. To quantify the collimation obtained by the ejected flow we plot the opening angle $\alpha$, given by $\alpha = \tan^{-1}\left[V_z^2 \left(V_x^2 + V_y^2\right)^{-1/2}\right]$. The part of the flow disrupted by the instabilities is located approximately between $z \sim 5 - 23$. Large variations in $T_e$, $V_z$ and $\rho$ are present, and the density drops below the vacuum cut-off at $z \sim 14$ mm. There is also no coherent collimation and the flow is sub-magnetosonic.

Instead at heights above $z \sim 23$ mm the emerging jet is found. Overall the jet is non-uniform with large density variations up to 2 decades in magnitude, but shows much smaller modulations in the axial velocity $V_z \sim 290 - 550$ km s$^{-1}$. The collimation attained by the flow is in fact very high ($\alpha < 20^o$ everywhere) with most regions collimated to within $10^o$; the fast mode Mach number of the flow is everywhere greater than two, with a maximum values of $M \sim 10$. The variation in axial velocity along the flow and its high collimation mean that fast moving regions of the jet may overtake slower propagating parts. The interaction may then develop into internal shocks in a mechanism similar to the internal shocks observe in YSO[57]. The plasma beta is considerably higher in the jet than in the surrounding cavity but it is still on average $\beta < 1$, however there are small "blobs" of plasma in the flow where $\beta \sim 1-5$.

Figure 8 also shows the ratio $B_\phi / |\mathbf{B}_p|$. The initially dominant toroidal component $B_\phi$ driving the magnetic tower is now approximately equal to the other components, $B_\phi \sim B_r \sim B_z$ corresponding to $B_\phi / |\mathbf{B}_p| = 1/\sqrt{2} \sim 0.7$, and the initial large-scale coherence present in the field does not exist any longer.

The redistribution of currents and fields, following the instability and detachment of the jet, is evident when considering the magnetic energy in the cavity. The initial current path is along the walls of the cavity and in the jet, with a very small fraction of the current flowing in the low density plasma in the cavity. As a simple approximation we can assume the magnetic tower to be an inductor consisting of two concentric, conducting cylinders of height $Z_t$; the outer cylinder corresponds to the cavity and has radius $R_c$, the inner cylinder with radius $R_j$ is the jet (see Figure 3). Currents flow along the surfaces of the two cylinders and connect radially at the height $Z_t$, the inductance is then given by $L = \frac{\mu_0}{2\pi} Z_t \ln\left(\frac{R_c}{R_j}\right)$. The values of $R_c$ and $Z_t$ are taken from the simulations and the jet radius is taken constant $R_j = 1$ mm. The inductance depends only weakly on the ratio $R_c/R_j$, and the constant jet radius assumption changes the results only slightly. The total magnetic energy inside the cavity is given by $E_M = \frac{1}{2} L I^2$, where $I$ is the current through the system; which is approximately constant (within 5%) between 205 and 255 ns. The MAGPIE generator has a high impendence and the current driven through the load does not significantly depend on the load's impedance. The magnetic energy $E_M$ is plotted in Figure 10, where it is compared to the actual total magnetic energy calculated in the simulations. There is good agreement until ~ 230 ns, indicating that the simple inductor picture and current distribution discussed above is essentially correct. However after ~ 230 ns, corresponding to the onset of large amplitude perturbations in the jet body, the total magnetic energy in the cavity does not increase as fast as $\frac{1}{2}LI^2$, but remains approximately constant. The currents redistribute in a way that is energetically favourable and minimize the magnetic energy in the cavity.

In general, the largest contribution to the total energy inside the magnetic tower is kinetic (Figure 10), while the internal energy (not shown in the Figure) is always < 1% of the total. The initial oscillations seen in the kinetic energy (for times less ~ 225 ns) are mainly due to current re-striking in the trailing plasma that is left by the non-uniform ablation and implosion of the wires. At later times however the relatively smaller variations are produced by the instabilities and break-up of the jet. For completeness, the kinetic energy, enthalpy and Poynting fluxes for the emerging jet are plotted as a function of axial position in Figure 8. The averaged fluxes are taken over a circular surface of radius $r = 3$ mm. The flow is dominated by the kinetic energy flux ($z \geq 23$ mm), while the enthalpy and Poynting fluxes are approximately equal.

Once it has completely emerged from the cavity, the jet propagates ballistically. The magnetic Reynolds number is only marginally higher than unity and the magnetic field transported with the jet flow is expected to diffuse over a time-scale of tens of ns, comparable to the dynamical timescale of the jet. However the heat generated through the field dissipation, should not destroy the collimation of the jet, which can still be efficiently cooled down by the radiation losses.

## IV. SUMMARY

We have presented the results of three-dimensional resistive MHD simulations of experiments designed to study the evolution of "magnetic tower" jets in the laboratory. The overall evolution of the laboratory jets shares a number of common physical processes with its astrophysical counterpart. The observed structure consists of a jet accelerated and confined by a toroidal field and embedded in a magnetic cavity. A shock envelope surrounds the magnetic cavity which elongates in time. The tower expansion is driven by the magnetic pressure and its velocity is well described by simply using the Rankine-Hugoniot relations across the shock at the head of the tower. However the growth of the tower is transient, the jet is unstable to current-driven modes which combined with the expansion through a steeply decreasing density background plasma result in the break-up of the structure. Magnetic fields and currents re-arrange and in place of the original coherent toroidal magnetic field a chaotic field develops inside the cavity. Finally, a well collimated (opening angle $<10^o$), radiatively cooled and clumpy jet is observed to emerge out of the cavity; the kinetically dominated jet is everywhere super-magnetosonic. Although there is currently no mechanism in the experimental set-up, the transient behaviour of the laboratory magnetic tower does not preclude the formation of a new tower and jet. By combining the burst of the cavity, which reorganizes the currents to flow close to the electrodes again, with a means to re-supply mass could trigger once more the formation of a new magnetic tower and the emergence of another clumpy jet.

**Acknowledgements**

The present work was supported in part by the European Community's Marie Curie Actions – Human Resource and Mobility within the JETSET (Jet Simulations

Experiments and Theory) network under contract MRTN-CT-2004 005592. The authors also wish to acknowledge the SFI/HEA Irish Centre for High-End Computing (ICHEC) and the London e-Science Centre (LESC) for the provision of computational facilities and support.# References

[1] Bruce A. Remington, R. Paul Drake, and Dmitri D. Ryutov, Reviews of Modern Physics 78, 755 (2006).
[2] D. R. Farley, K. G. Estabrook, S. G. Glendinning et al., Physical Review Letters 83, 1982 (1999).
[3] James M. Stone, Neal Turner, Kent Estabrook et al., Astrophysical Journal Supplement Series 127, 497 (2000).
[4] J. M. Foster, B. H. Wilde, P. A. Rosen et al., Physics of Plasmas 9, 2251 (2002).
[5] J. M. Foster, B. H. Wilde, P. A. Rosen et al., Astrophysical Journal 634, L77 (2005).
[6] A. Ciardi, S. V. Lebedev, J. P. Chittenden et al., Laser and Particle Beams 20 (2), 255 (2002).
[7] S. V. Lebedev, J. P. Chittenden, F. N. Beg et al., Astrophysical Journal 564 (1), 113 (2002).
[8] D. J. Ampleford, S. V. Lebedev, A. Ciardi et al., Astrophysics and Space Science 298, 241 (2005).
[9] S. V. Lebedev, D. Ampleford, A. Ciardi et al., Astrophysical Journal 616, 988 (2004).
[10] S. V. Lebedev, A. Ciardi, D. J. Ampleford et al., Plasma Physics and Controlled Fusion 47, 465 (2005).
[11] Mitchell C. Begelman, Roger D. Blandford, and Martin J. Rees, Reviews of Modern Physics 56, 255 (1984).
[12] Bo Reipurth and John Bally, Annual Review of Astronomy and Astrophysics 39, 403 (2001).
[13] Tsvi Piran, Reviews of Modern Physics 76, 1143 (2005).
[14] T. J. Galama, P. M. Vreeswijk, J. van Paradijs et al., Nature 395, 670 (1998).
[15] K. Z. Stanek, T. Matheson, P. M. Garnavich et al., Astrophysical Journal 591, L17 (2003).
[16] J. M. LeBlanc and J. R. Wilson, Astrophysical Journal 161, 541 (1970).
[17] A. M. Khokhlov, P. A. Höflich, E. S. Oran et al., Astrophysical Journal 524, L107 (1999).
20

**FIGURE CAPTIONS.**

FIGURE 1. (Colour online) Schematic of a radial wire array. The plasma background is formed as the ablated plasma from the wires is accelerated axially by the $\mathbf{j} \times \mathbf{B}_G$ force. The simulations are performed on a Cartesian (*xyz*) grid, however because of the symmetry of the radial array, we shall often describe vector components as axial (*z*-direction), radial (*r*-direction), toroidal ($\phi$-direction or azimuthal direction) or poloidal (vectors lying in the *r-z* plane).

FIGURE 2. (Colour online) (a) Time sequence of *x-z* slices of mass density. The times shown are, left to right, 180, 200, 220 and 240 ns respectively. The box size is $25 \times 40.7$ mm. The grey areas represent the electrodes (the central electrode has a radius of 2 mm and the outer electrode had a radius of 12.5 mm). The density range is logarithmic from $10^{-7}$ g cm$^{-3}$ (*blue*) to $10^{-2}$ g cm$^{-3}$ (*red*). The black line shows the density contour corresponding to $\rho = 10^{-4}$ g cm$^{-3}$, and serves as a guide when looking at (b) where an isodensity surface with $\rho = 10^{-4}$ g cm$^{-3}$ is shown at similar times.

FIGURE 3. (Colour online) Schematic of the magnetic tower. Various features observed and discussed in the text are shown The small red arrows indicate the current flow. Poynting flux through the base, injects magnetic energy in the tower.

FIGURE 4. (Colour online) Azimuthally averaged profiles at 220 ns. The radial profiles (a) are taken at $z = 8.7$ mm and at $z = 12.6$. The axial profiles (b) are taken at $r = 0$ mm corresponding to the axis and at $r = 2$ mm at the radius of the inner electrode.

FIGURE 5. (Colour online) (a) Synthetic x-ray emission images are shown from left to right at times 210, 220, 230 and 240 ns respectively. The intensity is on logarithmic scale and covers a range of $10^3$.
(b) Experimental, time-resolved XUV images taken at four different times (from left to right 268, 278, 288, 298 ns). Although the radial array used in this particular experiment has a different set-up from the one discussed in the simulations, the overall dynamical evolution is similar.

FIGURE 6. (Colour online) Axial profiles are obtained by azimuthally and radially averaging several quantities of interest. The profiles are shown at four different times (indicated at the top of the image).

FIGURE 7. (Colour online) Radial profiles are obtained by azimuthally and axially averaging several quantities of interest. The profiles are shown at four different times (indicated at the top of the image).

FIGURE 8. (Colour online) Launched jet properties. The plots show azimuthally ($\phi = 0\ldots 2\pi$) and radially averaged ($r = 0\ldots 1$ mm) profiles for several quantities of interest at 270 ns. The jet is found approximately at height $z > 23$ mm.

FIGURE 9. (Colour online) Magnetic field lines inside the cavity at 230 ns. The height of the box is 16 mm and the square base has 10 mm sides.

FIGURE 10. (Colour online) Total kinetic and magnetic energy in the magnetic tower obtained in the simulations. The energy values shown are for the inside of the cavity and do not include the shock envelope, which would mostly add to the kinetic and internal energies. The *blue* line is the calculated magnetic energy assuming the system is a simple inductor.

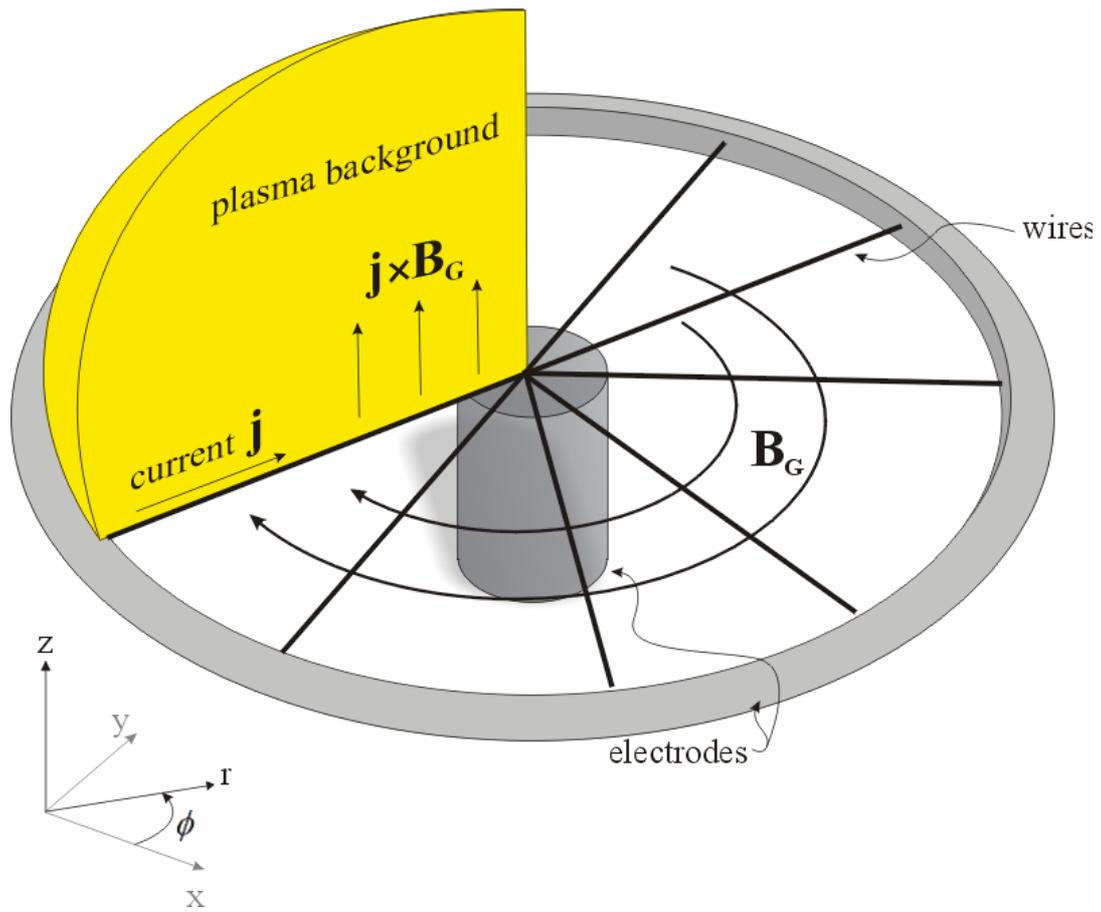

FIGURE 1

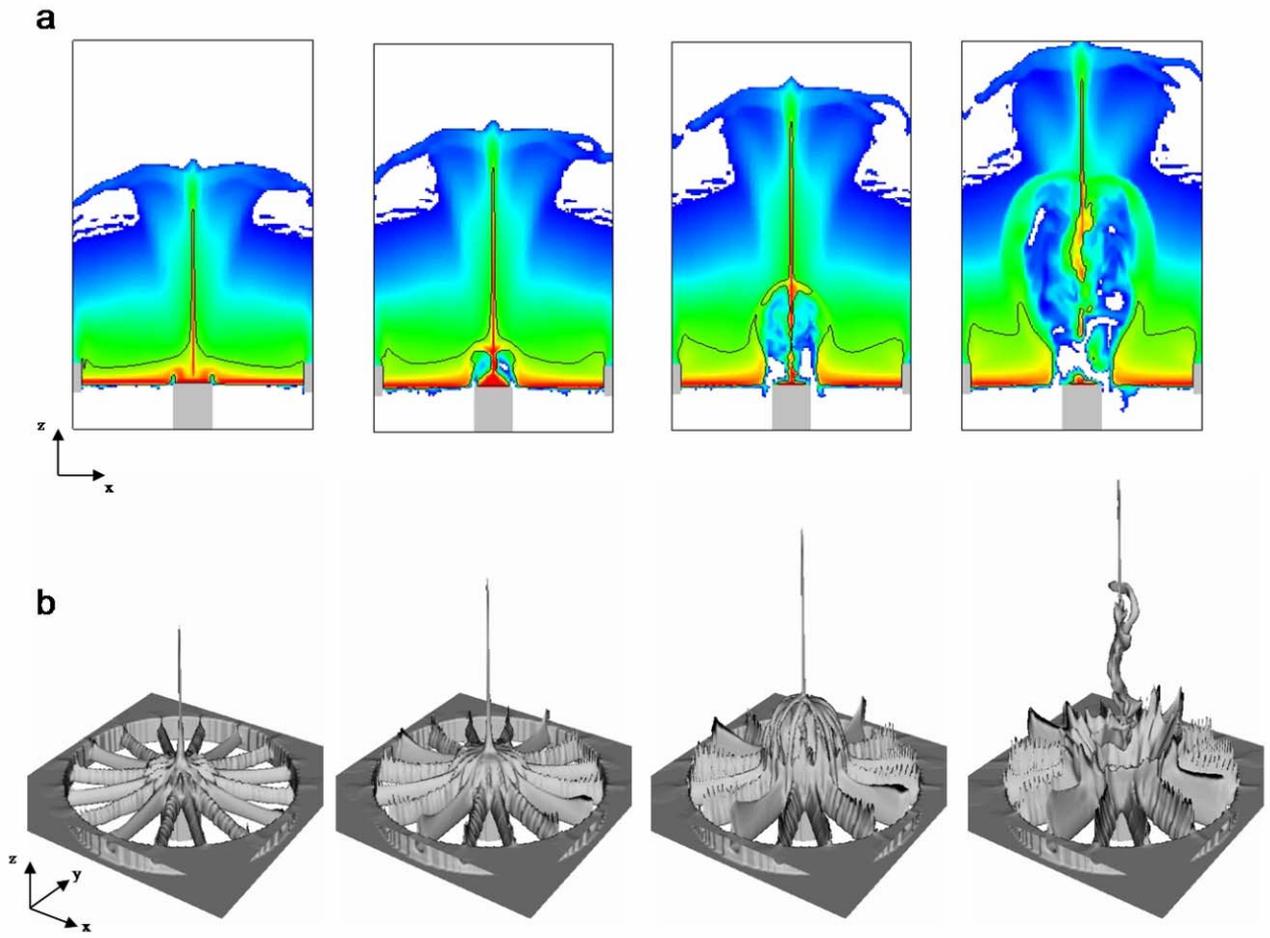

FIGURE 2

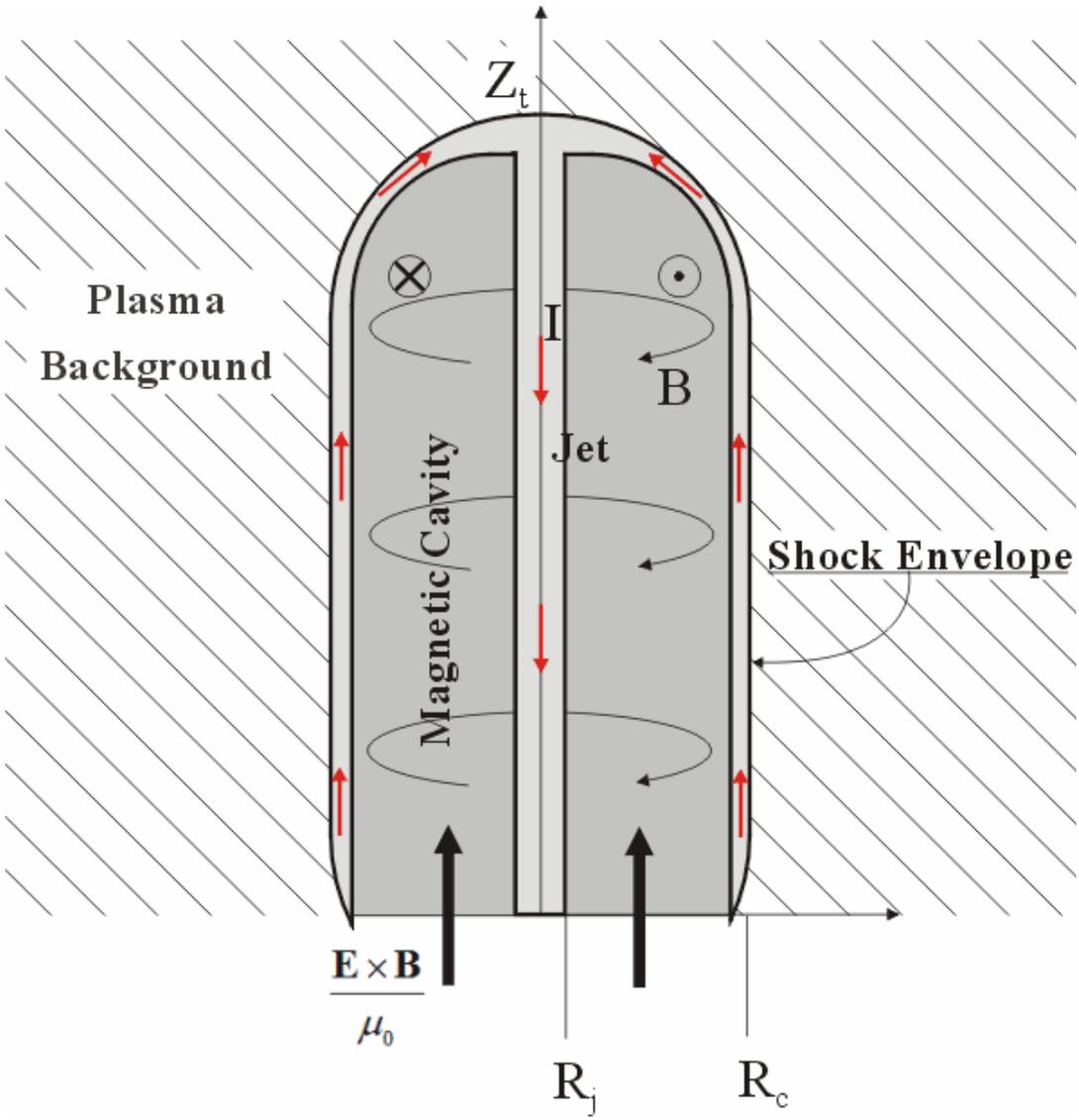

\

FIGURE 3

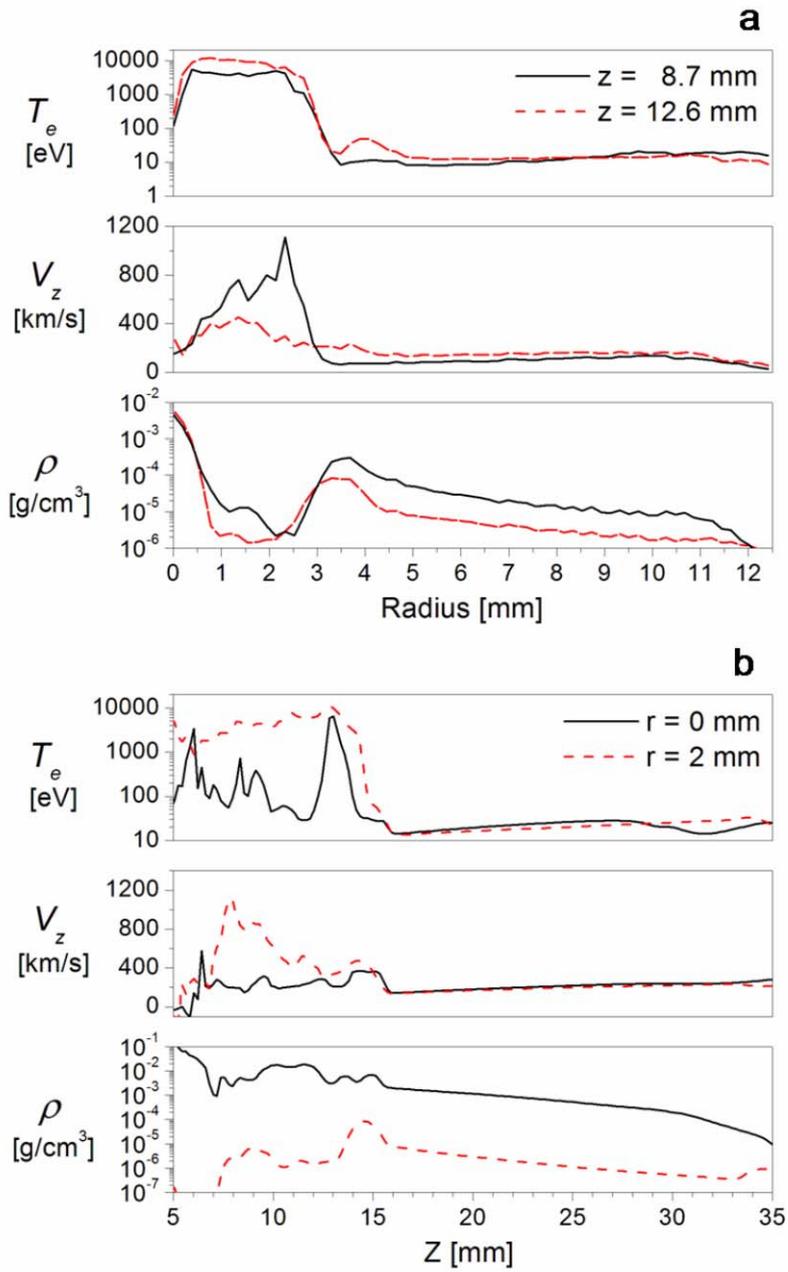

FIGURE 4

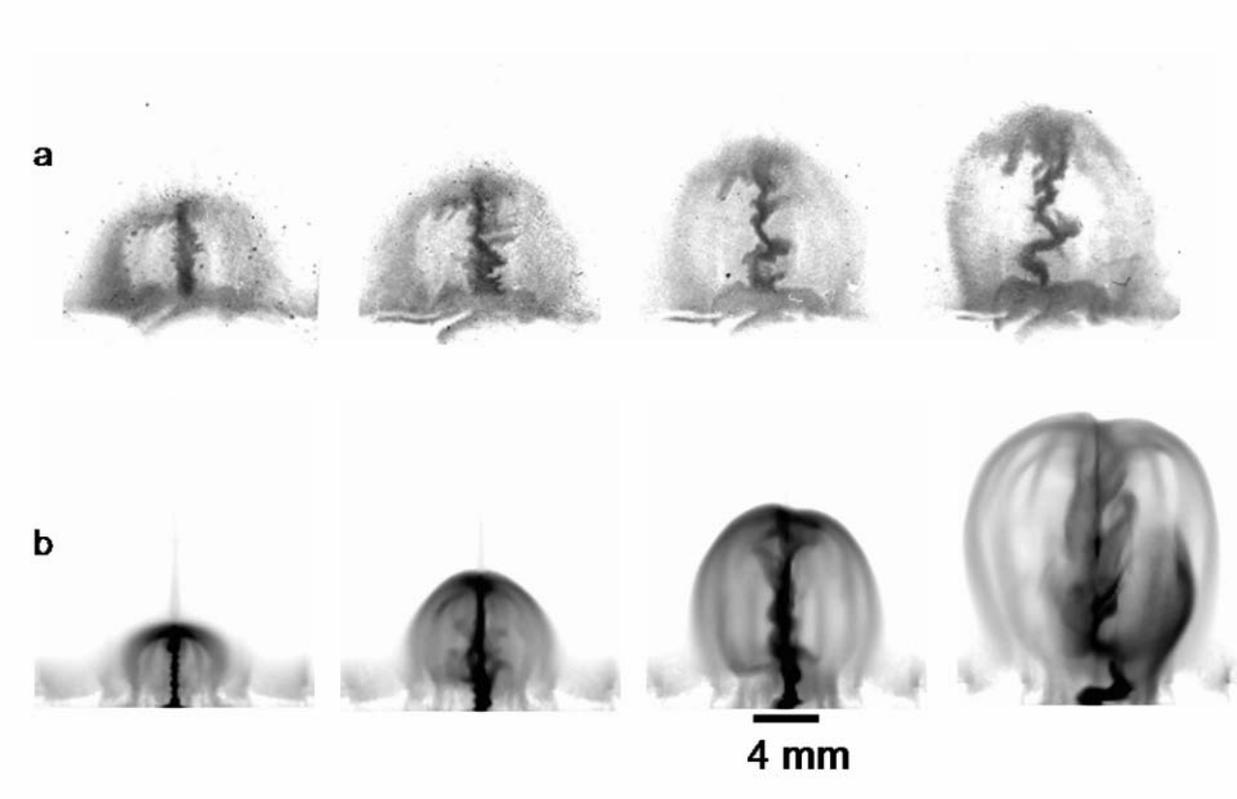

FIGURE 5

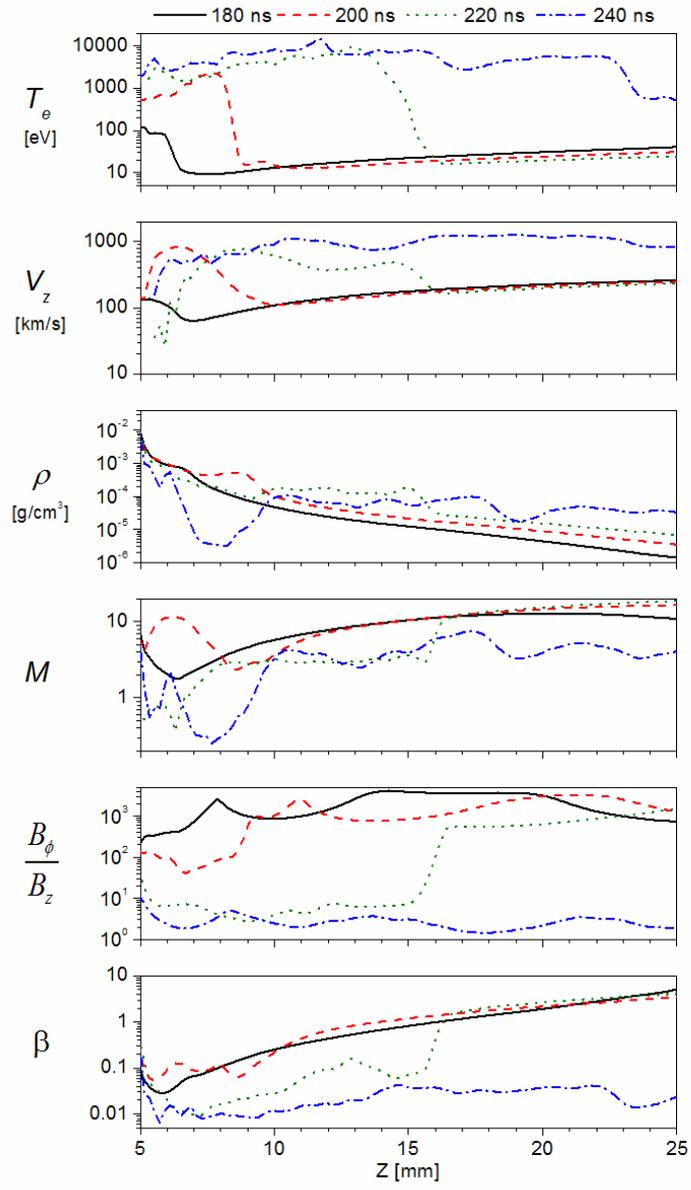

FIGURE 6

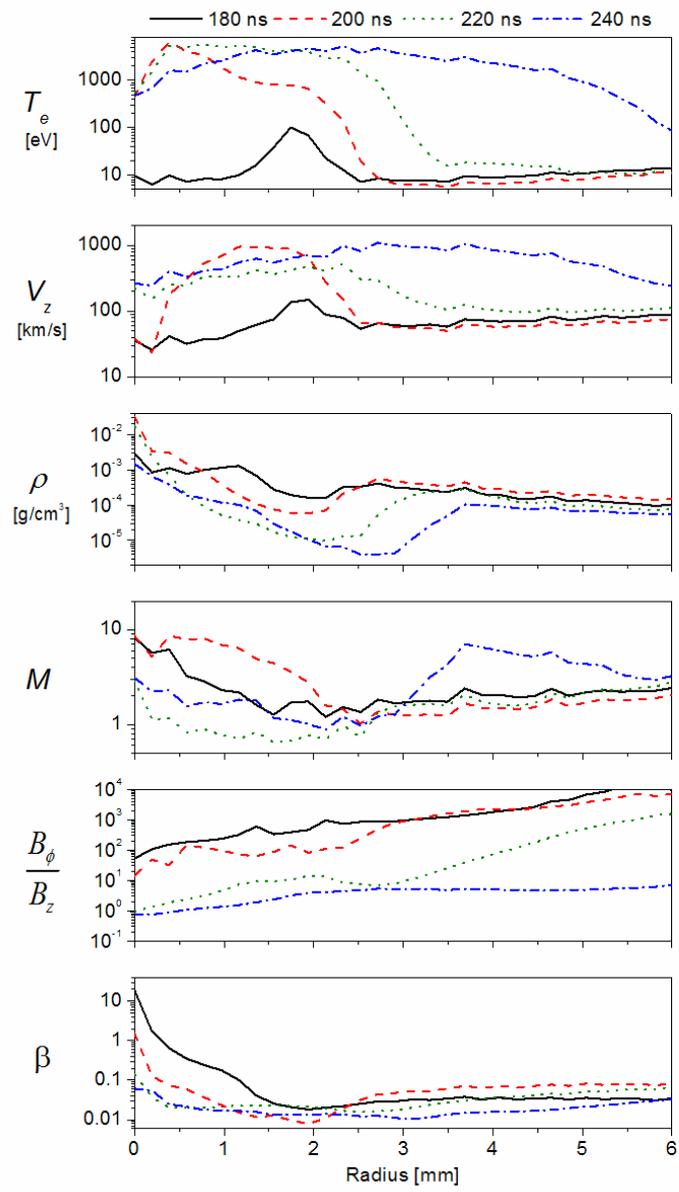

FIGURE 7

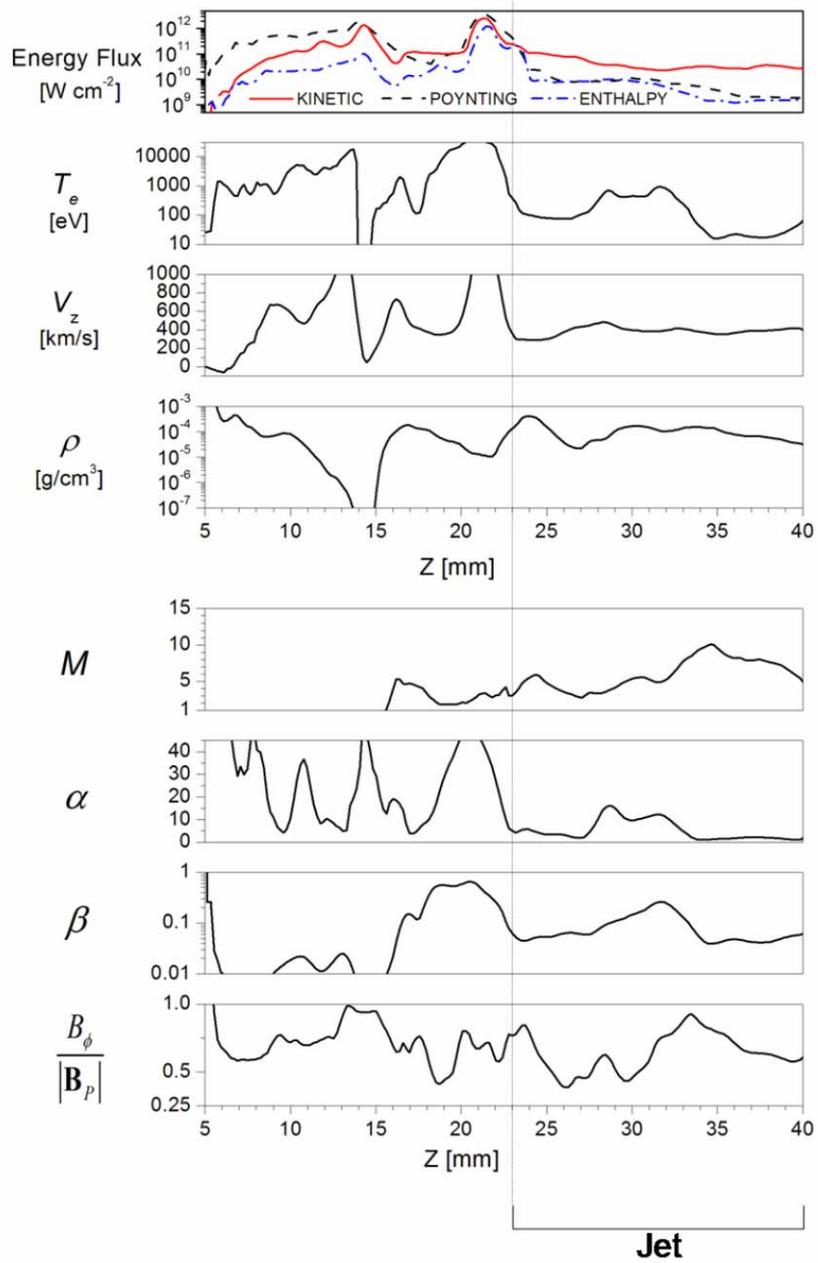

FIGURE 8

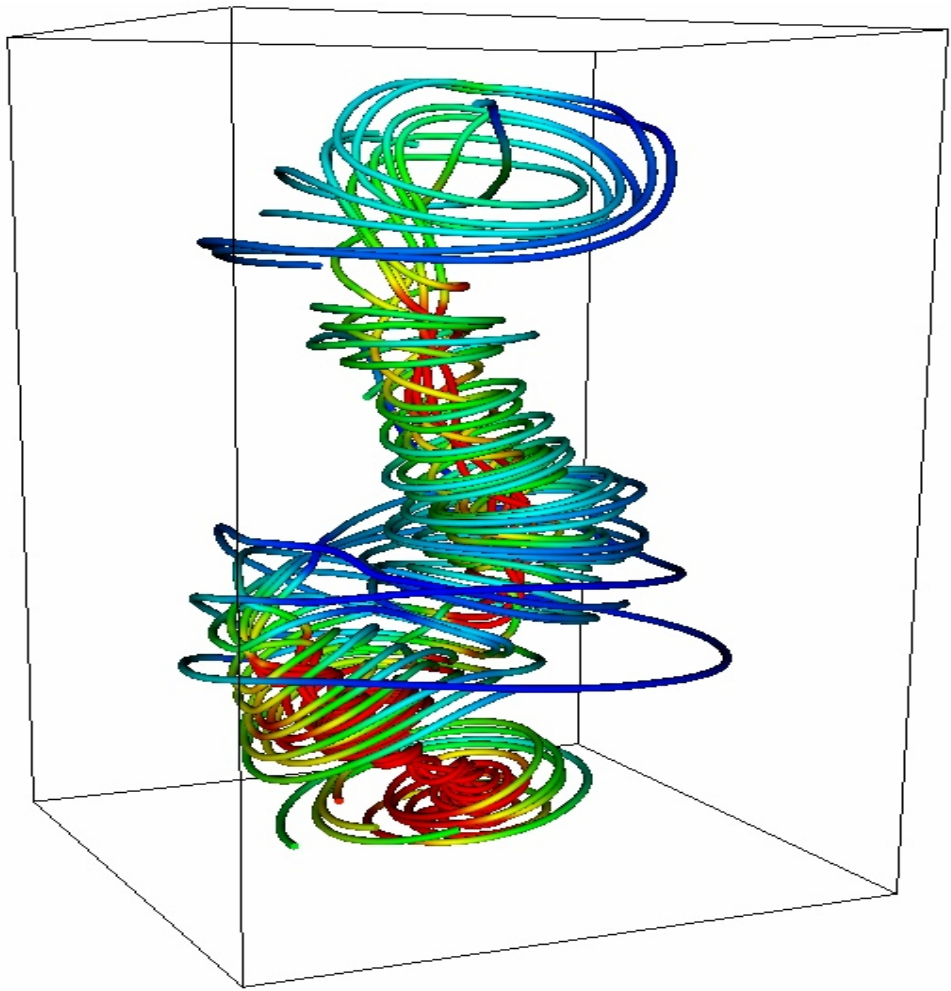

FIGURE 9

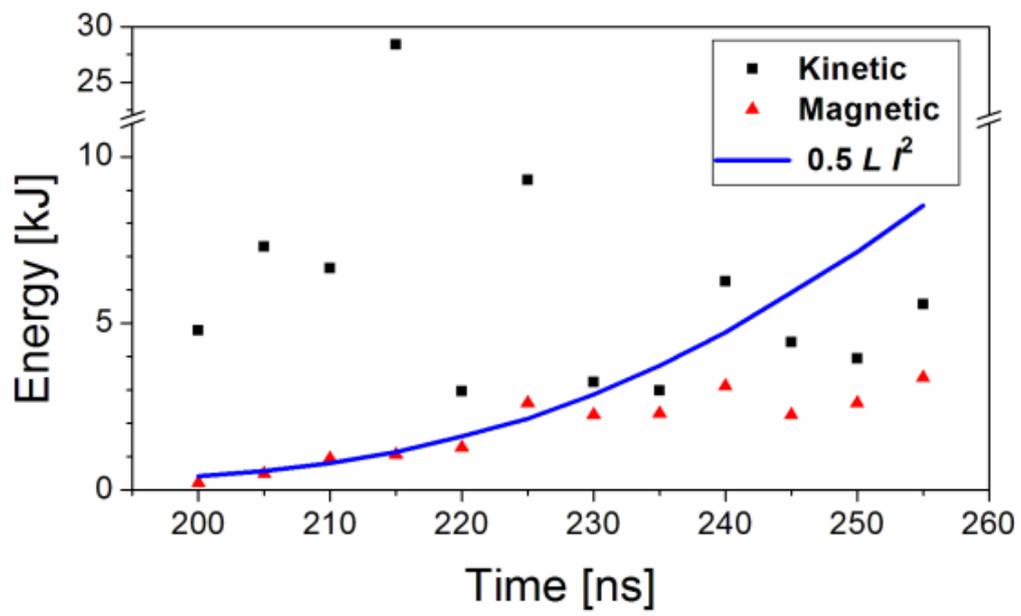

FIGURE 10